\documentclass[twocolumn,showpacs,preprintnumbers,amsmath,amssymb]{revtex4-1}

\usepackage{graphicx}
\usepackage{color}
\usepackage{bm}
\usepackage{bbm}
\usepackage{amsmath}
\usepackage{amsfonts}
\usepackage{subfigure}
\usepackage{overpic}
\usepackage{float}

\newcommand{\ket}[1]{\left| #1 \right\rangle} 
\newcommand{\bra}[1]{\left\langle #1 \right|} 
\newcommand{\braket}[2]{\left\langle #1 \vphantom{#2} \right|
 \left. #2 \vphantom{#1} \right\rangle} 

\begin{document}
\title{Quantum search on graphene lattices}

\author{Iain Foulger} \author{Sven Gnutzmann} \author{Gregor Tanner}
\affiliation{School of Mathematical Sciences, University of
  Nottingham, University Park, Nottingham NG7 2RD, UK.}

\date{\today}

\begin{abstract}
  We present a continuous-time quantum search algorithm on a graphene lattice.
  This provides the sought-after implementation of  an efficient
   continuous-time quantum search on a two-dimensional lattice.
  The search uses the linearity of the dispersion relation near the
  Dirac point and can find a marked site on a graphene lattice faster than
  the corresponding classical search.
 The algorithm can also be used for state transfer and communication.

\end{abstract}
\maketitle


{\em Introduction.-- }
Quantum walks \cite{Por13,Rei12} can provide polynomial and even exponential
speed-up compared to classical random walks 
\cite{ADZ93,kempe, Kendon06, Santha08}
and may serve as a
universal computational primitive for quantum computation 
\cite{Childs09}. This has led to
substantial interest in the theoretical aspects of this phenomenon, as well as in finding
experimental implementations \cite{Kar09,
  Sch09,Xue09,Zae10,Per08,Schr10}.  One of the most fascinating
applications of quantum walks is their use in spatial quantum search
algorithms first published for the search on the hypercube in \cite{SKW03}. 
Like Grover's search algorithm \cite{Gro96, Gro97} for
searching an unstructured database, quantum walk search algorithms 
can achieve up to quadratic speed-up compared to the corresponding
classical search. For quantum searches on $d$-dimensional square
lattices, certain restrictions have been observed, however, depending
on whether the underlying quantum walk is discrete \cite{ADZ93} or
continuous \cite{FG98}.  While effective search algorithms for discrete walks
have been reported for $d\ge 2$ \cite{AKR05, ADMP10}, continuous-time 
quantum search algorithms on square lattices show speed-up compared to 
the classical search only for $d\ge 4$ \cite{CG04a}.
This problem has been circumvented in \cite{CG04b}, however, at 
the conceptual cost  of adding internal degrees of
freedom (spin) and a discrete Dirac equation. 

Experimental implementations of discrete quantum walks
need time stepping
mechanisms such as laser pulses
\cite{Kar09,Sch09,Xue09,Zae10,Schr10}. It is thus in general 
simpler to consider experimental realizations with continuous time evolution. 
However, in the absence of internal degrees of freedom, 
no known search
algorithm on lattices exists up to now in the physically relevant
regime $d=2$ or $3$. Finding such an algorithm is highly topical
due to applications in secure state transfer and communication across
regular lattices as demonstrated in \cite{HT09I}.

We will show in the following that continuous-time quantum search in 2D is 
indeed possible! We will
demonstrate that such a quantum search can be performed at the Dirac 
point in graphene. This is
potentially of great interest, as graphene is now becoming available
cheaply and can be fabricated routinely \cite{CN09,WA11}. 
Performing quantum search and
quantum state transfer on graphene provides a new way of channeling
energy and information across lattices and between distinct sites.
Graphene sheets have been identified as a potential single-molecule 
sensor \cite{Sch07, Weh08} being very sensitive to a change of the 
density of states near the Dirac point. This property is closely related to 
the quantum search effect described in this paper.

Continuous-time quantum search algorithms take place on a lattice with
a set of $N$ sites interacting via hopping potentials
(usually between nearest neighbors only).  Standard searches work
at the ground state energy which, due to the periodicity of the
lattice, is related to quasi-momentum $k = 0$. After introducing a
perturbation at one of the lattice sites, the parameters are adjusted
such that an avoided crossing between the localized `defect' state and
the ground state is formed. The search is now performed in this
two-level sub-system \cite{HT10}. Criticality with respect to the dimension 
is reached when the gap at the avoided crossing and the eigenenergy 
spacing near the crossing scale in the same way with $N$.

Continuous-time quantum walks (CTQW) \cite{FG98} operate in
the position (site) space.  If the states $\ket{j}$ represent the
sites of the lattice, the Schr\"{o}dinger equation governing the
probability amplitudes $\alpha_{j}\left(t\right) =
\left<j\right|\left. \psi(t)\right>$ is given by
\begin{equation}\label{ctqw}
  \frac{d}{dt}\alpha_{j}\left(t\right) = -i\sum_{l=1}^{N} {H}_{jl}\alpha_{l}\left(t\right)
\end{equation}
where the Hamiltonian ${\bf H} = \epsilon_D {\bf I} + v {\bf A}$ is of
tight-binding type where 
$\bf A$ is the adjacency matrix of the lattice and $\bf I$ is the
identity matrix, $\epsilon_D$ is the on-site energy and $v$ is the
strength of the hopping potential. In \cite{CG04a}, the walk
Hamiltonian was set to be the discrete Laplacian where
$v = -1$ and $\epsilon_D$ is the coordination number of the lattice. 
A marked site  is then introduced by altering 
the on-site energy of that site. The
system is initialized at $t=0$ in the ground state of the unperturbed
lattice leading to an effective search for $d\ge 4$.
For the search based on the discretized Dirac operator \cite{CG04b}, 
an additional spin degree of freedom is introduced. 
This gives optimum search times for lattices with dimension
$d\geq3$ and a search time of $O(\sqrt{N}\ln N)$ for $d=2$ recovering
the results for discrete time walks \cite{AKR05}. 
We note that $d=2$ is the critical dimension in the discrete case
independent of the lattice structure; one thus finds an 
$O(\sqrt{N}\ln N)$ also for discrete time walks on
graphene \cite{ADMP10}.

The lack of speed-up for continuous search algorithms 
in two dimensions can be
overcome by making two adjustments: \textit{i.} the avoided crossing on
which the search operates is moved to a 
part of the spectrum
with a linear dispersion relation; \textit{ii.}
the local perturbation is altered in order to  
couple a localized perturber state and the
lattice state in the linear regime. 
The first point is addressed by
considering graphene lattices with the well-known linear dispersion
curves near the Dirac point. The perturbation at
the marked site is achieved by locally changing the hopping potential
 (instead of changing the on-site energy as in \cite{CG04a,CG04b}). We
start by giving an introductory account of basic properties of the
graphene lattice and its band structure \cite{CN09, WA11}.

\begin{figure}[t]
  \centering
  \includegraphics[width=0.45\linewidth]{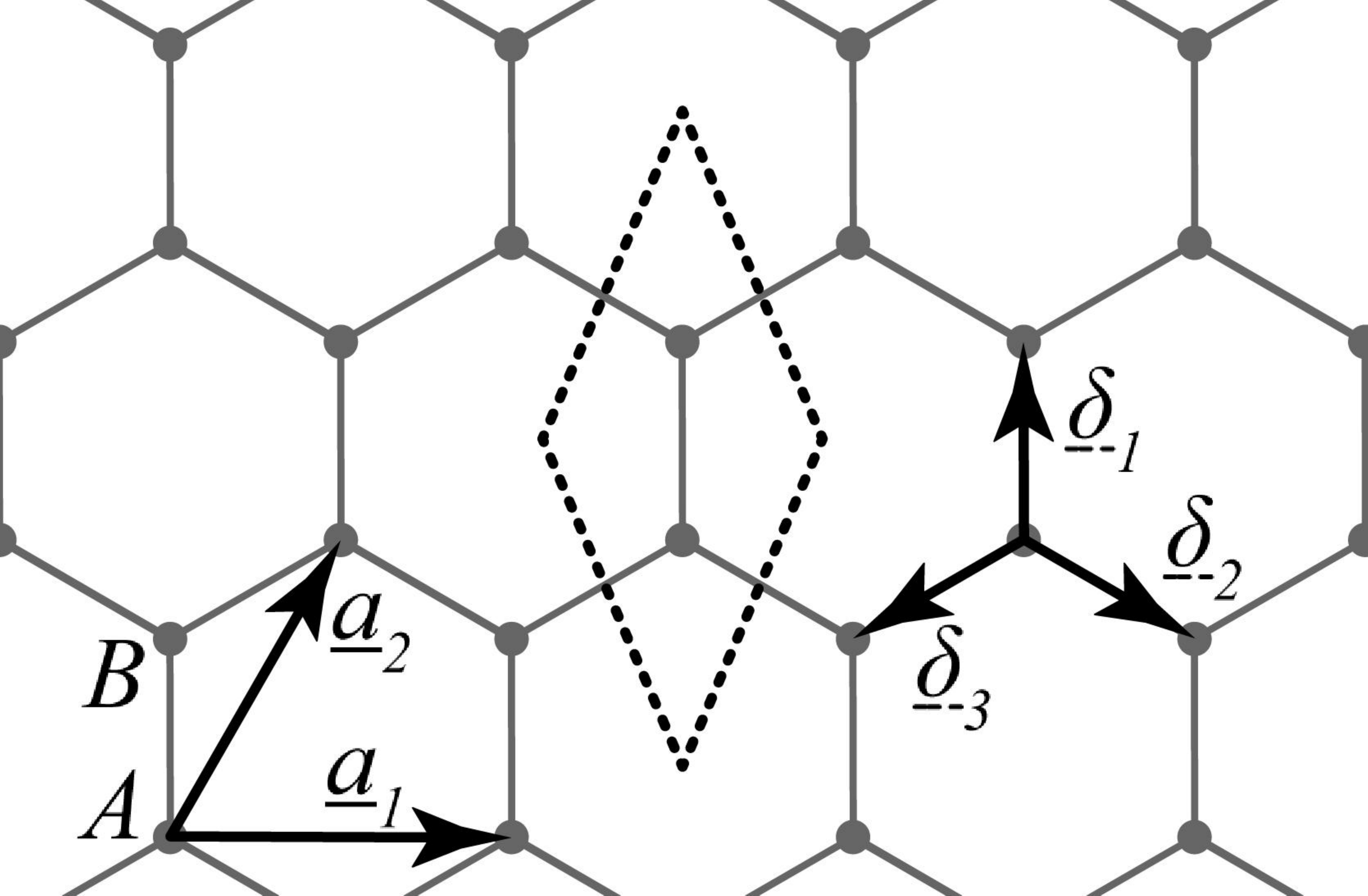}
   \hspace{.2cm}
  \includegraphics[width=0.45\linewidth]{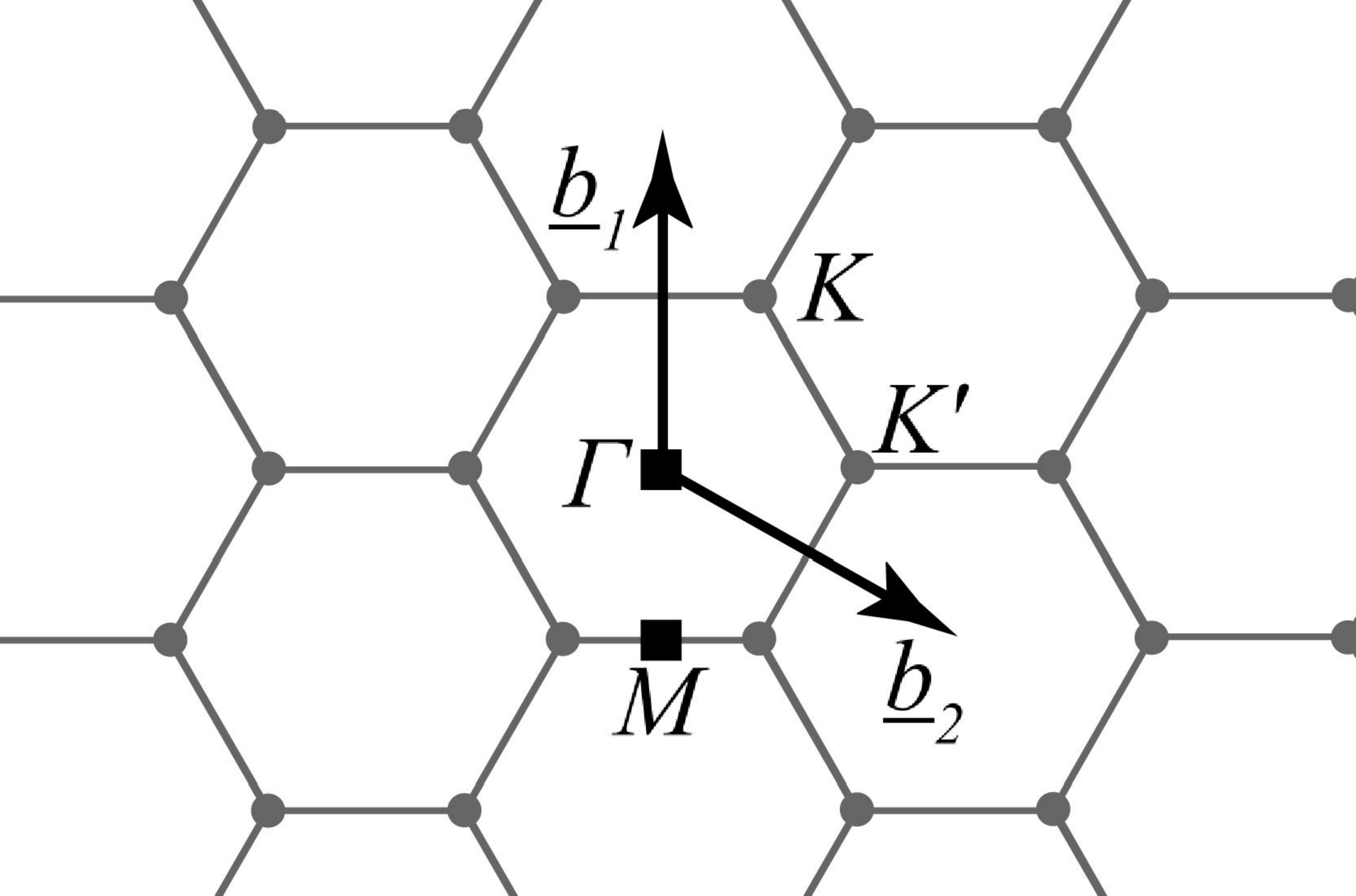}
  \label{fig:reciprocal_lattice}
   \caption{Left: Graphene with lattice vectors
    $\underline{a}_{1/2}$, 
    translation vectors
    $\underline{\delta}_{i}$
    and unit cell (dashed lines). Right: Reciprocal
    lattice with basis vectors
    $\underline{b}_{1/2}$, symmetry
    points
    $\underline{\Gamma}$, $\underline{K}$, $\underline{K}'$, $\underline{M}$
    and first Brillouin zone (hexagon).}
  \label{fig:both_lattices}
\end{figure}
{\em Review of graphene.-- }
The graphene (or honeycomb) lattice is bipartite 
with two
triangular sublattices, labeled A and B. 
The position of a cell in the lattice is denoted
by  $\underline{R} = \alpha \underline{a}_{1} + \beta \underline{a}_{2} $
where $\alpha$ and $\beta$ are integers and $\underline{a}_{1(2)}$
are basis vectors of the lattice (see
Fig.~\ref{fig:both_lattices}). States on the two sites within
one cell will be denoted by $|R\rangle^{A(B)}\equiv|\alpha,\beta\rangle^{A(B)}$.
The
corners of the Brillouin zone  (see
Fig.~\ref{fig:both_lattices})
are denoted $\underline{K}$ and
$\underline{K}'$ and the primitive cell contains two of these points.

\begin{figure}
  \centering
  \includegraphics[trim = 0mm 63mm 0mm 92mm,clip = true, width=0.9\linewidth]{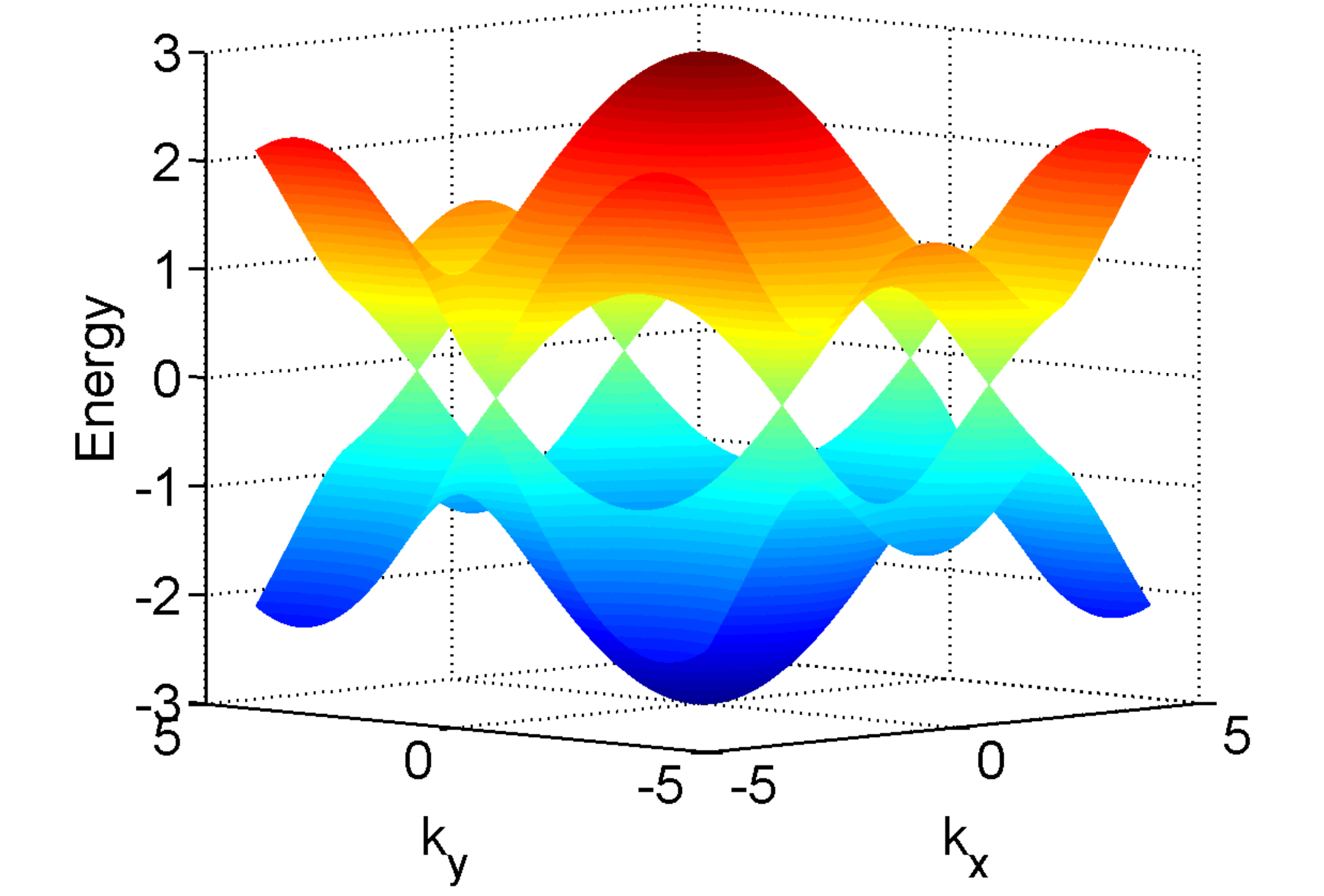}
  \caption{(Color online) Dispersion relation for infinite graphene sheet
    ($\epsilon_{D}=0$).}
  \label{fig:infinite_graphene_lattice}
\end{figure}

The solution for the tight-binding Hamiltonian 
on graphene as described above is well-known \cite{CN09, WA11}
and leads to the dispersion relation
\begin{eqnarray} \label{band} &&\epsilon\left(\underline{k}\right) =
  \epsilon_{D} \pm \\ \nonumber &&v\sqrt{1 +
    4\cos^{2}\left(\frac{k_{x}a}{2}\right) +
    4\cos\left(\frac{k_{x}a}{2}\right)\cos\left(\frac{\sqrt{3}k_{y}a}{2}\right)}
\end{eqnarray}
depicted in Fig.~\ref{fig:infinite_graphene_lattice} for an infinite
graphene lattice. 
It is indeed linear near the
Dirac points $\underline{K}\text{ and }\underline{K}'$ at the energy
$\epsilon_D$ where the conduction and valence
bands meet.
Around the Dirac points the dispersion relation $\epsilon(\underline{k})$ can be approximated by
\begin{equation}
  \epsilon\left(\underline{k}\right) \approx \epsilon_{D} \pm va\frac{\sqrt{3}}{2}\sqrt{\delta k_{x}^{2} + \delta k_{y}^{2}} = \epsilon_{D} \pm va\frac{\sqrt{3}}{2}\left|\delta k\right| .
\end{equation}

In the following, we will consider finite graphene lattices
with periodic boundary conditions, i.e. $|\Psi\rangle= 
\sum_{\alpha=1}^m\sum_{\beta=1}^n \left(\psi_{\alpha,\beta}^A |\alpha,\beta\rangle^A
+ \psi_{\alpha,\beta}^B |\alpha,\beta\rangle^B\right)$ with 
$\psi^{A(B)}_{\alpha,\beta}=\psi^{A(B)}_{\alpha+m,\beta}=\psi^{A(B)}_{\alpha,\beta+n}$. 
This simplifies the analysis allowing us to focus on the relevant features of the search
by avoiding boundary effects. The general description does not change for 
other boundary conditions, the localization amplitude on the marked site, however, 
becomes site dependent in a non-trivial way. Understanding this dependency is 
not essential in the context of this paper.

We denote $\underline{S}=
m\underline{a}_{1} + n\underline{a}_{2}$ the vector describing the
spatial dimensions of the lattice. 
Using Bloch's theorem \cite{Kit05}, the momentum is quantized as
\begin{equation} \label{k-xy}
  k_{x} = \frac{2\pi p}{ma} \hspace{5mm} k_{y} = \frac{1}{\sqrt{3}}\left(\frac{4\pi q}{na} - k_{x}\right)	
\end{equation}
where 
$p \in \left\{0,1,\ldots m-1\right\}, q \in \left\{0,1,\ldots n-1\right\} $ 
and the spectrum
(\ref{band}) becomes discrete. In what follows, for simplicity, we have assumed that our lattice is square
in the number of cells, that is, that $m=n=\sqrt{\frac{N}{2}}$.
Four-fold degenerate states with energy $\epsilon_{D}$ and wave numbers exactly on the 
Dirac points $\underline{K} \text{ and } \underline{K}'$
exist if $m$ and
$n$ are some multiples of 3. We assume this in the following
for simplicity. In the general case,
one needs to consider the 
states closest to the Dirac energy which gives a more complex 
theoretical analysis while 
essential signatures do not change. 

{\em Quantum search.-- }
Setting up a continuous-time search  
by changing the on-site energy of the marked
site  as done in  \cite{CG04a} 
does not work for graphene.  Using the ground state as the
starting state fails for the same reason as it fails for rectangular
lattices in $d=2$ or $3$ as the dispersion relation is quadratic near
the ground state, see Fig.~\ref{fig:infinite_graphene_lattice}. 
Alternatively, moving the search to the Dirac point implies
constructing an avoided crossing between a localized perturber 
state and a Dirac state. As the Dirac energy coincides with 
the on-site energy $\epsilon_{D}$, this leads to the condition, that
the on-site energy perturbation must vanish at the crossing, which 
brings us back to the unperturbed lattice.

We therefore mark a given site by changing the
hopping potentials between the site and its nearest neighbors. 
Focusing on a symmetric choice of the perturbation 
and setting $\epsilon_{D}=0$ for 
convenience, we obtain
the (search-) Hamiltonian
\begin{equation}\label{eq:graphene_search_hamiltonian} {\bf H}_\gamma
  = -\gamma {\bf A} + {\bf W}.
\end{equation} 
Here, ${\bf W}$ denotes the perturbation changing the hopping potential to and from the marked site
$(\alpha_0,\beta_0)^A$ which has been chosen to be on the $A$ lattice, that is,
\begin{equation}\label{eq:W}{\bf W} =
  \sqrt{3}\ket{\alpha_0,\beta_0}^{A} \bra{\ell}
  + \sqrt{3}\ket{\ell} \bra{\alpha_0,\beta_0}^{A}.
\end{equation}
The state $\ket{\ell}$ denotes the symmetric superposition of the three neighbors 
of the marked site, that is, 
\begin{equation}
  \ket{\ell}=\frac{1}{\sqrt{3}}\left(\ket{\alpha_0,\beta_0}^{B}+\ket{\alpha_0,\beta_0-1}^{B}
    + \ket{\alpha_0+1,\beta_0-1}^{B}\right).
    \label{localizedstate}
\end{equation}

At $\gamma = 1$, the perturbation corresponds to a
hopping potential $v=0$ between the site $(\alpha_0,\beta_0)^A$ and its
neighbors, effectively removing the site from the lattice. It is this
perturbation strength which is important in the following.
Experimentally, such a perturbation is similar to graphene lattices 
with atomic vacancies as they occur naturally in the production 
process \cite{Mey08}; in microwave analogs of graphene as 
discussed in \cite{Kuhl10}, this can be realized by removing single
sites from the lattice.

\begin{figure}
\begin{overpic}[width = 1.05\linewidth]{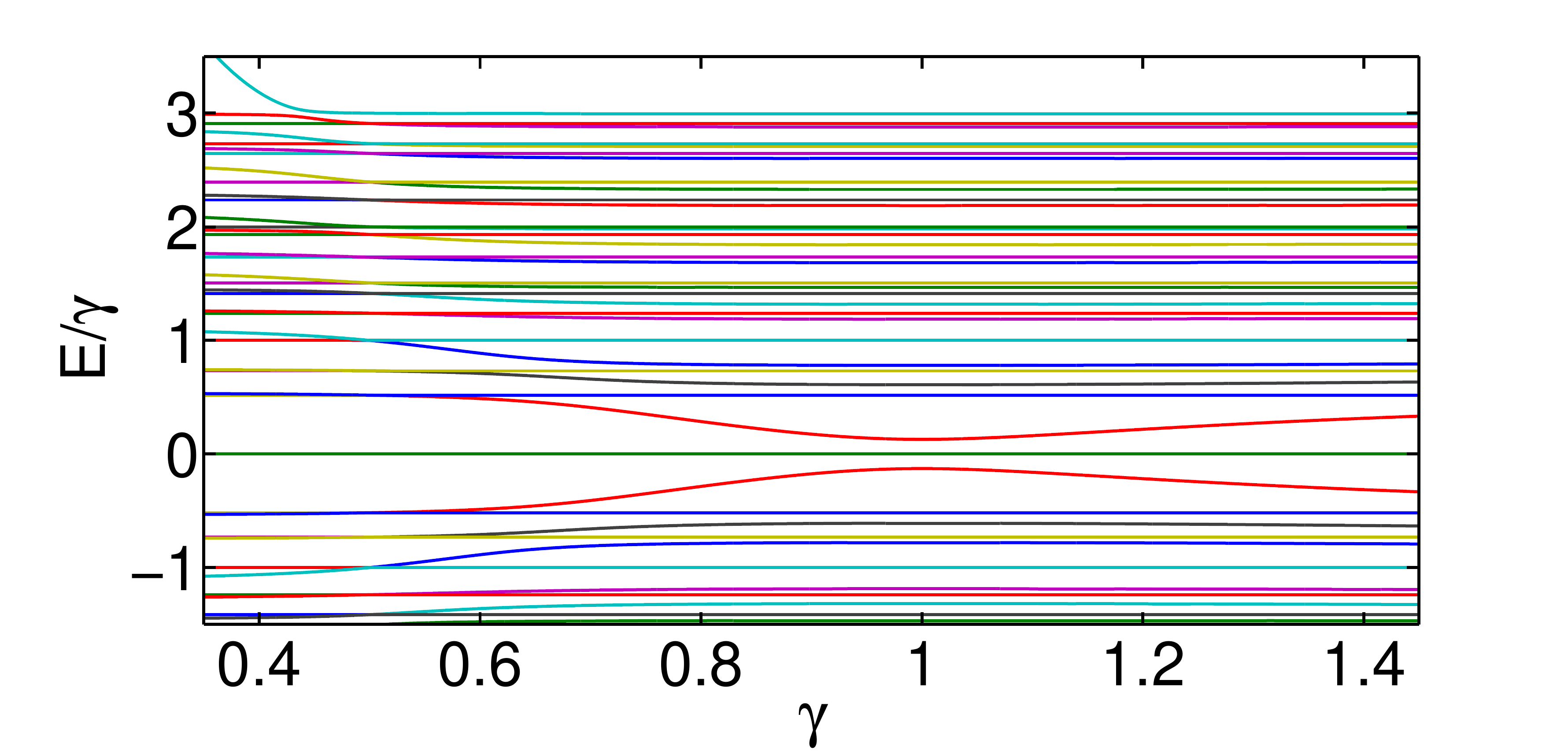}
        \put(54.25,26.75){\fbox{\includegraphics[width=0.37\linewidth]{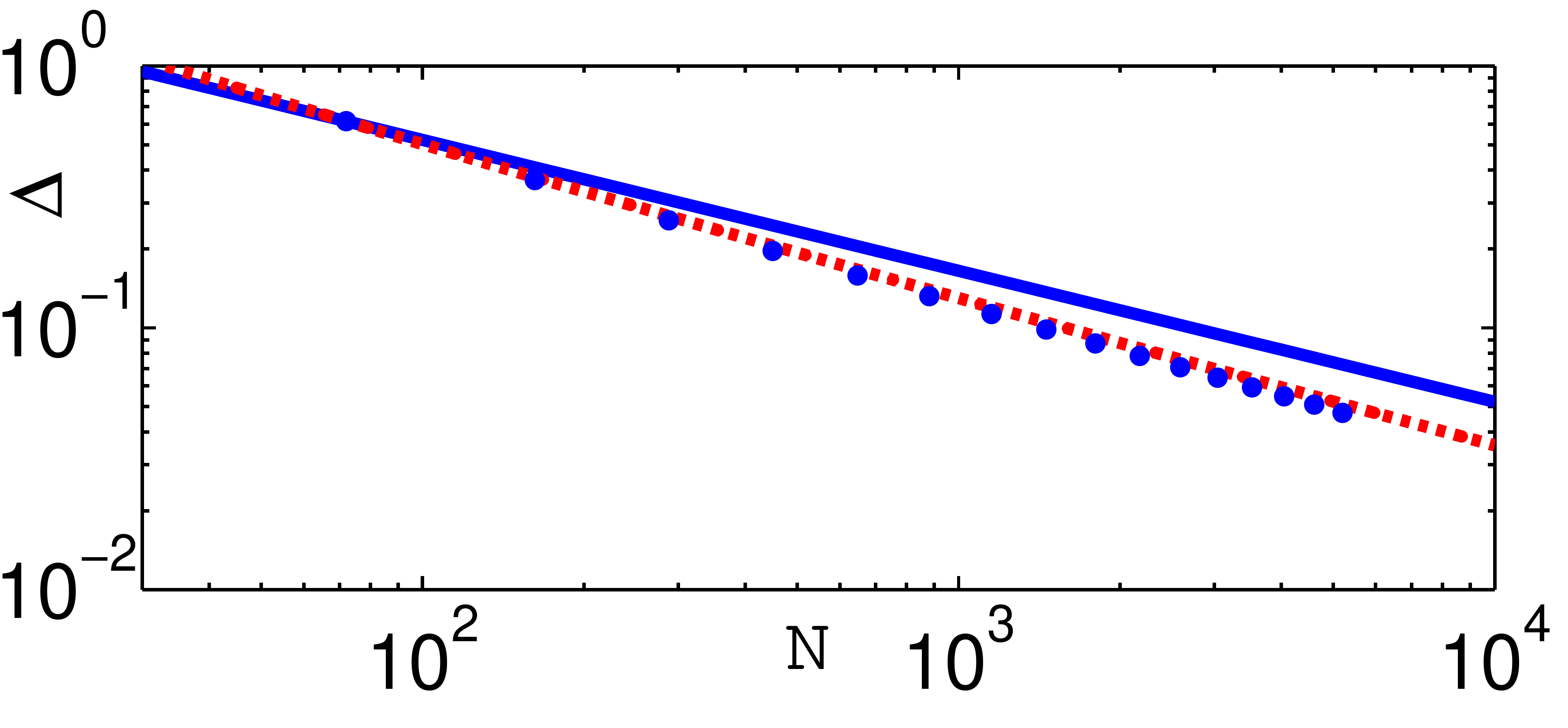}}}
\end{overpic} 
 \caption{Spectrum $ {\bf H}_\gamma$ in Eq.~\ref{eq:graphene_search_hamiltonian}
    as a function of $\gamma$ for a $12\times 12$ cell torus  ($N=288$). The
    spectrum is symmetric around $\epsilon_{D} = 0$.  Inset: Scaling
    of the gap $\Delta = \tilde{E}_{+} - \tilde{E}_{-}$ (dots) and curves $c_{1}/\sqrt{N}$ 
    (solid blue), $c_{2}/\sqrt{N\log N}$ (dashed red) for comparison. }
  \label{fig:3bond_spectrum}
\end{figure}

The effect of marking (or perturbing) the graphene Hamiltonian can be
seen numerically in the parametric behavior of the spectrum of ${\bf H}_\gamma$ 
as a function of $\gamma$, see Fig.~\ref{fig:3bond_spectrum}
for the case $n=m=12$. Note that $\bf W$ is a rank two perturbation
which creates two perturber states. 
These states start to interact with the
spectrum of the unperturbed graphene lattice from $\gamma \approx 0.5$
onwards working their way through to a central avoided crossing at
$\gamma = 1, E = 0$. Below we will show, how the avoided crossing can 
be used for searching;  note, that the parameter dependence of the avoided 
crossing ($\gamma=1$) is evident from the tight-binding Hamiltonian
$H$ in (\ref{eq:graphene_search_hamiltonian}). In a realistic set-up,  
the perturbation needs to be fine-tuned in general to be in resonance with 
an eigenstate of the (unperturbed) system near the Dirac point.     

At the avoided crossing there are altogether six states 
close to the Dirac energy:
the two perturber states and the four degenerate Dirac
states
\begin{align}\label{eq:K_state}
  \ket{K}^{A(B)} =&
  \sqrt{\frac{2}{N}}
  \sum_{\alpha, \beta} e^{i\frac{2\pi}{3}\left(\alpha + 2\beta + 2\sigma \right)}
  \ket{\alpha,\beta}^{A(B)}\nonumber \\
  \ket{K'}^{A(B)} =&\sqrt{\frac{2}{N}}
   \sum_{\alpha, \beta} e^{i\frac{2\pi}{3}\left(2\alpha + \beta\right)}\ket{\alpha,\beta}^{A(B)}
\end{align}
where $\sigma=1$ ($\sigma=0$) for states on the B (A) lattice and $N=2n m$ is the number of sites in the lattice. 
One finds directly ${\bf W}
\ket{K}^B = {\bf W}
\ket{K'}^B=  0$, that is, Dirac states on the
B lattice do not interact with an A-type perturbation 
for all $\gamma$.
Furthermore, at $\gamma= 1$, the marked
state $\ket{\alpha_0,\beta_0}^A$ is an eigenvector of 
${\bf H}_{\gamma=1}$
with eigenvalue $E=0$ - the marked site is disconnected from
the lattice. 

Thus, the avoided crossing involves only the two Dirac states
$\ket{K}^A$, $\ket{K'}^A$ and one perturber state
$\ket{\tilde{\ell}}$.
Neglecting the interaction of the perturbation 
with the rest of the spectrum at the avoided crossing,
we set
$\ket{\tilde{\ell}}\approx \ket{\ell}$ (see \eqref{localizedstate})
and use this to reduce the full
Hamiltonian locally in terms of the $3$-dimensional basis 
$\{\ket{K}^A, \ket{K'}^A, \ket{\ell}\}$.
The reduced Hamiltonian takes the form
\begin{equation}\label{redH}
  \tilde{\bf H} = \sqrt{\frac{6}{N}} \begin{bmatrix}
    0 & 0 & e^{-i\frac{2\pi}{3}\left(\alpha_{o} + 2\beta_{o}\right)}  \\
    0 & 0 & e^{-i\frac{2\pi}{3}\left(2\alpha_{o} + \beta_{o}\right)} \\
    e^{i\frac{2\pi}{3}\left(\alpha_{o} + 2\beta_{o}\right)} & e^{i\frac{2\pi}{3}\left(2\alpha_{o} + \beta_{o}\right)} & 0   
  \end{bmatrix}
\end{equation}
with eigenvalues $ \tilde{E}_{\pm} = \pm 2\sqrt{\frac{3}{N}}, \tilde{E}_{0}
= 0$, and eigenvectors
\begin{align} \label{eq:reduced_matrix_eigenvectors}
  \ket{\tilde{\psi}_{\pm}} &= \frac{1}{2}\left(e^{-i\frac{2\pi}{3}\left(\alpha_{o} + 
  2\beta_{o}\right)}\ket{K}^A + e^{-i\frac{2\pi}{3}\left(2\alpha_{o} + \beta_{o}\right)}\ket{K'}^A \pm \sqrt{2}\ket{\ell}\right) \\
  \ket{\tilde{\psi}_{0}} &=
  \frac{1}{\sqrt{2}}\left(e^{-i\frac{2\pi}{3}\left(\alpha_{o} +
        2\beta_{o}\right)}\ket{K}^A -
    e^{-i\frac{2\pi}{3}\left(2\alpha_{o} +
        \beta_{o}\right)}\ket{K'}^A \right).
\end{align}
For searching the marked site $(\alpha_0,\beta_0)^A$, 
the system is initialized in a delocalized starting state
involving a superposition of Dirac states. This state will then rotate
into a state localized on the neighbors of the marked site. 
The search is initialized in the optimal starting state
\begin{eqnarray}\label{eq:start_state}
  \ket{s} &=& \frac{1}{\sqrt{2}}\left(\ket{\tilde{\psi}_{+}}+\ket{\tilde{\psi}_{-}}\right) \\ \nonumber
  &=& \frac{e^{-i\frac{2\pi}{3}\left(\alpha_0 + 2\beta_0\right)}}{\sqrt{2}}
  \left(\ket{K}^A + e^{-i\frac{2\pi}{3}\left(\alpha_0-\beta_0\right)}\ket{K'}^A\right)
\end{eqnarray}
which still depends on the perturbed site. Lack of knowledge of $\left(\alpha_0, \beta_0\right)$
leads, however, only to an $N$ independent overhead,  see the discussion below. Letting
$\ket{s}$ evolve in time with  the reduced Hamiltonian (\ref{redH})
we obtain
\begin{align}
  \ket{\psi\left(t\right)} &= e^{-i\tilde{\bf H}t}\ket{s}
  = \frac{1}{\sqrt{2}}{\left(e^{-i\tilde{E}_{+}t}\ket{\tilde{\psi}_{+}} + e^{-i\tilde{E}_{-}t}\ket{\tilde{\psi}_{-}}\right)} \nonumber\\
  &{= \cos\left(\tilde{E}_{+}t\right)\ket{s} -
  i\sin\left(\tilde{E}_{+}t\right)\ket{\ell}},
\end{align}
that is, the system rotates from $\ket{s}$ to $\ket{\ell}$ in time $t = \frac{\pi}{4}\sqrt{\frac{N}{3}}$.
We find a $\sqrt{N}$ speed-up for the search on graphene. 
This, together with the linear dispersion relation near the Dirac point, 
where the spacing between successive eigenenergies scales like 
$O(1/\sqrt{N})$, makes the search on this 2D lattice possible.
In contrast to the algorithms described in 
\cite{CG04a,CG04b}, the system localizes here on the
neighbors of the marked site, the marked site can be found by three
additional direct queries. Furthermore, the initial starting state is
not the uniform state here, but the state $\ket{s}$ in
(\ref{eq:start_state}). To construct this initial state uniquely
requires some information about the site that is being searched for.
Without this knowledge, one has three possible optimal initial states for an
A-type perturbation as can be seen from
Eqn.~(\ref{eq:start_state}). 
The same applies for marking a B-type
site, so in total there are six possible optimal starting states. 
As these states are not orthogonal, this
increases the number of runs for a successful search by a factor of 4.
The additional overhead is independent of N, and thus does not alter the scaling 
with system size. In an experiment one may have little control about how the system is 
excited at the Dirac energy, so the initial state will be in a 
more or less arbitrary superposition of all four Dirac states. 
The search is then not optimal but runs with a success probability that is, 
on average, again reduced by a factor $1/4$.

\begin{figure}[t]
  \centering
  \includegraphics[width= 1.0\linewidth]{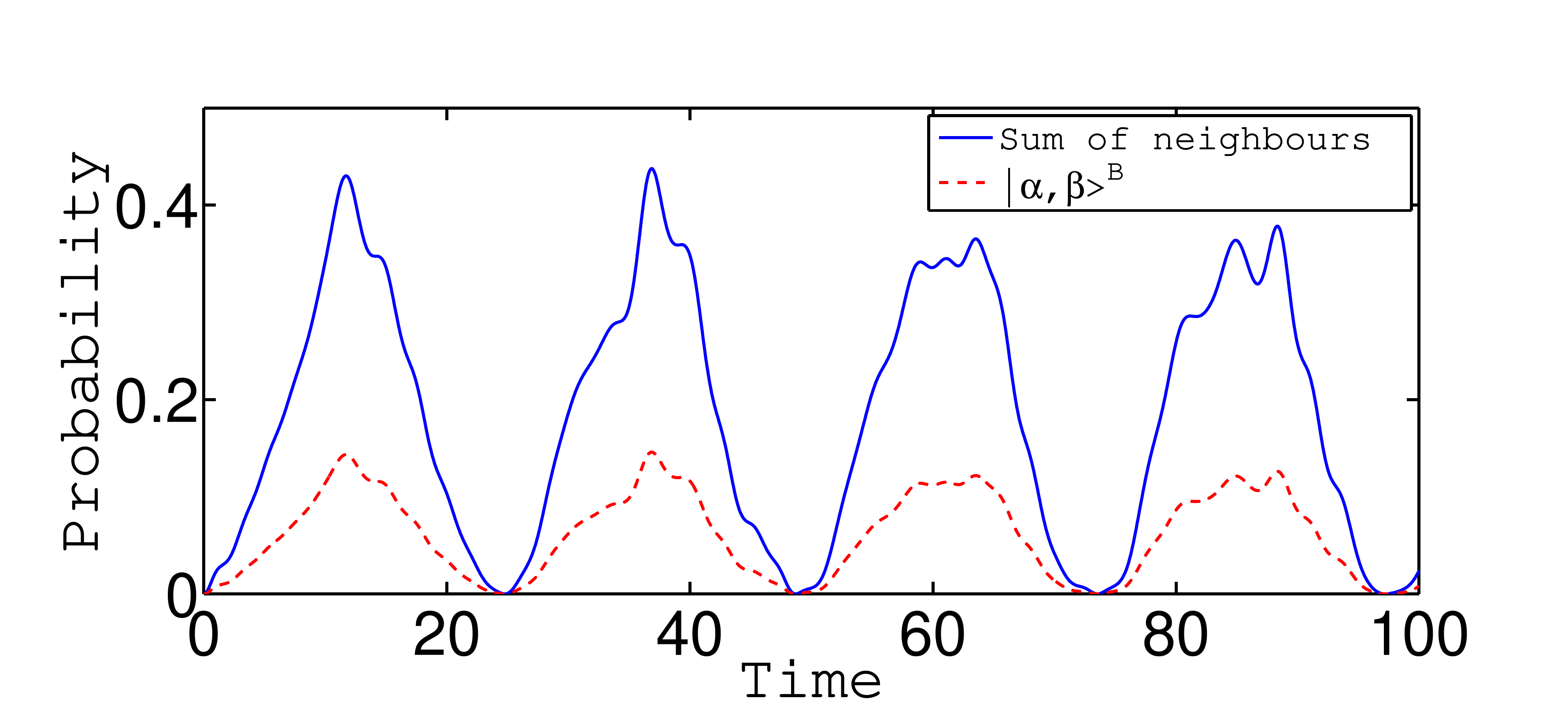}
  \caption{Search on $12\times 12$ cell graphene lattice with starting state $\ket{s}$. For tori with $m = n$ the 
    dynamics at each neighboring site is the same so only one is shown.}
  \label{fig:search_prob}
\end{figure}

Fig.~\ref{fig:search_prob} shows a numerically 
obtained quantum search initialized in $\ket{s}$ and
evolving under the full search Hamiltonian. As expected from the 
analysis on the reduced Hamiltonian, the state localizes on the three 
neighboring sites with a probability of about 45\% which is two orders of 
magnitude larger than the
average probability $100/N$, here roughly 0.5\%. The search does not
reach 100\% due to the fact that the actual localized state 
$|\tilde{\ell}\rangle$ extends beyond the nearest neighbors of 
the marked site, so $\langle{\tilde{\ell}}|{\ell}\rangle = O(1) <1$.

Our reduced model neglects contributions
from the rest of the spectrum; like for other 
discrete and continuous time walks at the critical dimension  
\cite{FG98, AKR05,ADMP10,CG04a,CG04b}, 
these contributions give $\ln N$ corrections (such as the 
$O\left(1/\sqrt{N\ln\left(N\right)}\right)$ scaling of the gap at
 the avoided crossing shown in the inset of
Fig.~\ref{fig:3bond_spectrum}). These logarithmic corrections have been 
derived in the appendix 
 by going beyond the reduced three-state model, 
see also  \cite{CG04a}.
The relevant exact eigenenergies $E_+ = -E_-$ at the avoided crossing satisfy the
resolvent condition 
\begin{equation}
F\left(E_{\pm}\right) =  \frac{\sqrt{3}}{N}
    \sum_{\underline{k}}
    \left[
      \frac{1}{E_{\pm} - \epsilon\left(\underline{k}\right)}
      +
      \frac{1}{E_{\pm} + \epsilon\left(\underline{k}\right)}
    \right] = 0\, ,
\end{equation}
with $\epsilon(k) >0$ the eigenenergies of the unperturbed system at quasi-momenta $k$  
given in (\ref{k-xy}). Expanding $F(E_+)=4\sqrt{3}/(NE_+)-\sum_{n=1}^\infty I_{2n}
E_{+}^{2n-1}$,
one finds $I_2= O(\mathrm{ln} N)$
and $I_{2k}=O(N^{k-1})$ for $k\ge 2$, 
see the appendix 
 for details. The scaling of
the gap follows then directly. 
The localization time scales inversely proportional to the
gap, that is, $T = O\left(\sqrt{N\ln\left(N\right)}\right)$; one also obtains
that the return amplitude drops like $O\left(1/\sqrt{\ln N}\right)$.

We note in passing that our search algorithm can - like all quantum searches - be 
used for quantum communication and state transfer. Following \cite{HT09I},  our continuous 
time search can be used to send signals between different sites by
adding an additional perturbation to the lattice. The quantum system is then initialized 
in a state localized on one of the perturbed sites and the system oscillates between states 
localized on the perturbations. We find that the mechanism works best when both
 perturbations are on the same sublattice. Due to the nature of the coupling
between the A and B sublattice and the fact that the localized perturber states live
(mostly) on one sublattice, signal propagation between perturbations on different 
sublattices  takes place over a much longer timescale.

{\em Discussion.-- } Continuous-time quantum search can be
performed {effectively}
on a 2D lattice without internal degrees of freedom by
running the search at the Dirac point in graphene. We find that our
search succeeds in time $T =
  O\left(\sqrt{N\ln\left(N\right)}\right)$ with probability $O\left(1/\ln N\right)$. This is the same time complexity found in \cite{AKR05,ADMP10} 
for discrete-time and in
\cite{CG04b} for continuous-time searches. 
To boost the probability to $O\left(1\right)$, $O\left(\ln N\right)$ 
repetitions are required 
giving a total time $T = O\left(\sqrt{N}\ln^{\frac{3}{2}}N\right)$.  
Amplification methods \cite{Gro97a, Tul08, PGKJ09} may
be used to reduce the total search time further.

For simplicity of the analysis, we have focused here on perturbations 
which alter the hopping potential to all three nearest-neighbors symmetrically. 
Efficient search algorithms can also be obtained  using other types of 
perturbations such as a single-bond
perturbation or perturbing the lattice by adding additional sites.  
In all cases, it is important to fine-tune the system parameters in order
to operate at an avoided crossing near the Dirac point. Given the
importance of graphene as a nano-material, our
findings point towards applications in directed 
signal transfer, state reconstruction or sensitive switching. This opens
up the possibility of a  completely new type of electronic engineering 
using single atoms as building blocks of electronic devices.     

\acknowledgements
This work has been supported by the EPSRC network `Analysis on Graphs'
(EP/I038217/1). Helpful discussions with Klaus Richter are gratefully acknowledged.

\newpage
\appendix
\section{Supplementary notes - Quantum search on graphene lattices}

 We present here supplementary notes related to the 
 article {\em Quantum search on graphene lattices}.
  There, we proposed an implementation of  an efficient
   continuous-time quantum search on a two-dimensional 
   graphene lattice.
  The notes here provide additional  details - not presented in the main text - on the scaling relation of the 
  search time and the success probability.
  We derive in particular the $\log N$ correction terms. 

The search dynamics is defined through a tight-binding Hamiltonian  $H$ consisting of the 
Hamiltonian of the unperturbed (finite) graphene lattice, $H_0$, and a perturbation term changing 
the hopping potential between a single {\em marked} site $\ket{\alpha_o,\beta_o}$ and its 
three neighbors. The perturbation has the effect of decoupling the marked site from its 
neighbors. The Hamiltonian is of the form 
\begin{equation}
H = H_0 + \sqrt{3}\ket{\alpha_o,\beta_o}\bra{\ell} + \sqrt{3}\ket{\ell}\bra{\alpha_o,\beta_o}
\label{eq:search_hamiltonian}
\end{equation}
and we assume periodic boundary conditions on the unperturbed graphene Hamiltonian 
$H_0$, see the main text for details.
Here, $\ket{\ell}$ is a uniform superposition of the three neighboring vertices adjacent to $\ket{\alpha_o,\beta_o}$. 
We derive in what follows the search time $t=T$, that is, the time at which the amplitude at $\ket{\ell}$  
reaches a maximum. The amplitude squared is interpreted as the success probability
for the search to succeed. The amplitude is determined by evaluating
\begin{equation}
  \bra{\ell}e^{-iHT}\ket{start} = \sum_{\ket{\psi_{a}} 
  } \braket{\ell}{\psi_{a}}\braket{\psi_a}{start}e^{-iE_{a}T},
  \label{eq:amplitude}
\end{equation}
with $\ket{\psi_a}$, $E_a$, the eigenstates and eigenenergies of the perturbed lattice.  
Note that $\ket{\alpha_o,\beta_o}$ is itself an eigenstate
with $H \ket{\alpha_o,\beta_o}=0$ and $\braket{\ell}{\alpha_o,\beta_o}=0$,
so that it does not contribute to the sum \eqref{eq:amplitude}.
Without loss of generality, we choose our initial state on the 
A sublattice of graphene, that is
\begin{equation}
  \ket{start} = \frac{1}{\sqrt{2}}\left(\ket{A_{\underline{K}}} + \ket{A_{\underline{K}'}}\right)
\end{equation}
with $\ket{A_{\underline{K}}},  \ket{A_{\underline{K}'}}$ being degenerate 
eigenenergies of the unperturbed lattice at the Dirac energy and on the 
A-lattice. (If the perturbation is on the B-lattice, the search is not 
successful and we repeat the search on the B sublattice). In these notes, we 
will justify our simple reduced model in the main text showing that our algorithm 
effectively takes place in a two-dimensional subspace of the full 
Hilbert-space spanned by combinations of the energy-states near the Dirac 
point and a localized perturber state. Our analysis follows the treatment in 
Ref.\ [S1] adjusted to  include symmetry properties of graphene around the 
Dirac point.

For any eigenstate $\ket{\psi_{a}}$
of the perturbed system such that the corresponding eigenenergy 
$E_a$ is not in the spectrum of $H_0$ we may rewrite
the perturbed eigenequation in the form
\begin{equation}
\ket{\psi_{a}} = \frac{\sqrt{3R_{a}}}{E_{a} - H_{0}}\ket{\alpha_{o},\beta_{o}},
\label{eq:perturbed_eigenvector_definition}
\end{equation}
where $\sqrt{R_{a}} = \braket{\ell}{\psi_{a}}$ 
(where we chosen the phase of $\ket{\psi_{a}}$
such that$ \braket{\ell}{\psi_{a}}\ge 0$).\\
If $E_a$ is in the spectrum of $H_0$ then
$\sqrt{R_a}=0$. This is trivial for the state  
$\ket{\psi_{a}}=\ket{\alpha_{o},\beta_{o}}$ -- otherwise it can be shown 
as follows. Let $\ket{\psi_a^0}$
be an unperturbed eigenvector such that 
$H_0 \ket{\psi_a^0}= E_a \ket{\psi_a^0}$.
Projecting the eigenvalue equation $H \ket{\psi_{a}}= E_a \ket{\psi_{a}}$
onto $\ket{\psi_a^0}$ yields 
$\braket{\psi_a^0}{\ell}\braket{\alpha_{o},\beta_{o}}{\psi_a}+
\braket{\psi_a^0}{\alpha_{o},\beta_{o}}\braket{\ell}{\psi_a}=0$.
Clearly, $\braket{\alpha_{o},\beta_{o}}{\psi_a}=0$; in addition for
a given $E_a$, we can always find at least one corresponding 
$\ket{\psi_a^0}$ such that
 $\braket{\psi_a^0}{\alpha_{o},\beta_{o}}\neq 0$. From that it follows
that $\braket{\ell}{\psi_a}=0 \equiv R_a^2$, which is what we wanted 
to show. Note that 
due to high degeneracies in the unperturbed system 
and due to the fact, that the perturbation is of rank two,
 one will have many states whose 
eigenenergies do not change under the perturbation and for which 
\eqref{eq:perturbed_eigenvector_definition} is a priori not well defined.
The property $R_a=0$ allows us to remove these states
from all sums that involve  $(E_a-H_0)^{-1}$.

Let us now use \eqref{eq:perturbed_eigenvector_definition}
to derive a condition for an energy $E_a\neq 0$ to be a \textit{perturbed}
eigenenergy of $H$ -- perturbed eigenenergy here implies that
it is not in the spectrum of $H_0$.
As $\ket{\alpha_{o},\beta_{o}}$ is a known eigenstate of the perturbed
lattice,
we have $\braket{\alpha_{o},\beta_{o}}{\psi_a} = 0$ and thus
\eqref{eq:perturbed_eigenvector_definition} implies
\begin{equation}
  \sqrt{3R_{a}}\bra{\alpha_{o},\beta_{o}}\left(E_{a} - H_{0}\right)^{-1}\ket{\alpha_{o},\beta_{o}} = 0\,.
\end{equation}
Expressing $\bra{\alpha_0,\beta_0}$ in terms of the eigenstates of the 
{\em unperturbed }
Hamiltonian $H_0$, we may write this as a quantization condition 
\begin{equation} \label{FE}
  \begin{split}
    F\left(E_{a}\right) =& 0\\
    F\left(E\right) =&  \frac{\sqrt{3}}{N}
    \sum_{\underline{k}}
    \left[
      \frac{1}{E - \epsilon\left(\underline{k}\right)}
      +
      \frac{1}{E + \epsilon\left(\underline{k}\right)}
    \right].
  \end{split}
\end{equation}
Here, $N$, is the total  number of sites and 
$\epsilon\left(\underline{k}\right)$ are the positive eigenenergies of the 
unperturbed Hamiltonian $H_0$. (Note that the spectrum of $H_0$ 
as well as $H$ is symmetric around $E=0$. This constitutes the main difference
to the treatment considered in [S1]. )\\
We may choose $\ket{\psi_{a}}$ to be normalized
$\braket{\psi_{a}}{\psi_{a}}=1$ -- \eqref{eq:perturbed_eigenvector_definition}
then implies
\begin{equation}
  3 R_{a}\bra{\alpha_{o},\beta_{o}}\left(E_{a} - H_{0}\right)^{-2}
  \ket{\alpha_{o},\beta_{o}} = 1,
\end{equation}
which allows $R_{a}$ to be rewritten as 
\begin{equation}
  R_{a} = \frac{1}{\sqrt{3}|{F'\left(E_{a}\right)}|}.
  \label{eq:Ra}
\end{equation}
We may now rewrite the
amplitude \eqref{eq:amplitude} in the form
\begin{equation}
  \begin{split}
    \bra{\ell}e^{-iHT}\ket{start} =&
      \sum_{a: R_a \neq 0} \sqrt{R_a} \braket{\psi_a}{start} e^{-iE_{a}T}\\
    =&
    \braket{\alpha_{o},\beta_{o}}{start}\sum_{a}\frac{e^{-iE_{a}T}}{E_{a}|{F'\left(E_{a}\right)}|}\, 
  \end{split}
  \label{eq:time_evolution}
\end{equation}
where we have used the adjoint of \eqref{eq:perturbed_eigenvector_definition}
and have removed the restrictions on the summation in the last line.
This is no longer necessary as $|F'(E_a)|\to \infty$ when $E_a$ is in the 
unperturbed spectrum.
In the main text, we show that by adding
a perturbation which 
creates 
a localized state energetically close to the Dirac point 
one can construct an efficient search algorithm.  
Consequently, we concentrate on evaluating the time-evolution 
involving the eigenstates of the perturbed 
Hamiltonian closest to the Dirac point; we denote these states $\ket{\psi_{\pm}}$ in what follows. We  estimate the corresponding eigenenergies, 
$E_{\pm}$ 
where $E_{+}=-E_{-}>0$ and we will focus on $E_+$ in the following.
Using the sum in Eq.\ (\ref{FE}), we will also derive a leading 
order expression for $F'\left(E_{+}\right)$.
Separating out the contribution to $F\left(E_{+}\right)$ from the Dirac 
points where $\epsilon\left(\underline{K}\right)= \epsilon\left(\underline{K}'\right) = 0$, and
expanding the remaining contribution to the sum in (\ref{FE}) at $E=0$, one obtains 
\begin{equation}\label{exp}
  F\left(E_{+}\right) = \frac{4\sqrt{3}}{NE_{+}} - \sum_{n=1}^{\infty} I_{2n}E_{+}^{2n-1}.
\end{equation}
The sums $I_{n}$ are given by
\begin{equation}
  I_{n} = \frac{\sqrt{3}}{N}\sum_{\underline{k} \neq \underline{K},\underline{K}'} 
    \left[
      \frac{1}{\left[\epsilon\left(\underline{k}\right)\right]^{n}}
      + 
      \frac{1}{\left[-\epsilon\left(\underline{k}\right)\right]^{n}}
    \right]
  \,. 
\end{equation}
Due to the symmetry of the unperturbed spectrum only those $I_{n}$ with 
even $n$ are non-zero.  

The non-vanishing $I_{2k }$ coefficients 
obey the following rigorous estimates
\begin{align}
  I_2=&  O \left(\ln N\right),
  \label{I2est}\\
  \lim_{N \to \infty } \frac{I_{2k}}{N^{k-1}}=&
  4 \sqrt{3} \left(Z_2(S_{\underline{K}},k)+ Z_2(S_{\underline{K}'},k)
  \right) \nonumber\\
  & \qquad \qquad \qquad \qquad 
  \text{for $k\ge 2$,}
  \label{Ikest}
\end{align}
where the estimate \eqref{I2est} 
is sharp ($I_2$ is logarithmically bounded from above and 
below) and 
$Z_2(S,x)$ is the Epstein zeta-function 
\begin{equation}
  Z_2(S,x)
  = \frac{1}{2} 
  \sum_{ (p,q) \in \mathbb{Z}^2 
    \backslash (0,0)} 
  \left( S_{11}p^2 
    +2 S_{12} pq + S_{22} q^2
  \right)^{-x}
\end{equation}
for a real positive definite real symmetric $2 \times 2$ matrix
$S$ [S2]. The matrices  
\begin{equation}
  S_{\underline{K}}=  S_{\underline{K}'}=
  4\pi^2 \begin{pmatrix}
    2 & -1 \\
    -1 & 2
  \end{pmatrix} 
\end{equation}
describe the spectrum close to the Dirac points. The linear dispersion behaviour near the
Dirac points $K$ and $K'$ is the same and so are the matrices $S_{\underline{K}}$
and $S_{\underline{K}'}$.\\
Before moving on let us derive the estimates \eqref{I2est}
and \eqref{Ikest}. 
It is clear that the dominant contributions come from
the vicinity of the Dirac points. Approximating the spectrum
close to the Dirac points one has
\begin{equation}
  \begin{split}
    I_{2k}=& 2\sqrt{3} N^{k-1} \left[ 
      \sum_{(p,q)\in L}
      \frac{1}{\left(S_{\underline{K}, 11}p^2 + 2S_{\underline{K}, 12}pq+
        S_{\underline{K}, 22}q^2 \right)^{k}}
      +\right. \\
    &\left.
      \sum_{(p,q)\in L}
      \frac{1}{\left(S_{\underline{K'}, 11}p^2 + 2S_{\underline{K'}, 12}pq+
        S_{\underline{K'}, 22}q^2 \right)^{k}}
    \right] + O(1) \ .
  \end{split}
\end{equation}
Here the sums over integers $p$ and $q$ is over a rectangular
region $L$ of the lattice $\mathbb{Z}^2$ which is centered at $(0,0)$
and has side lengths proportional to $\sqrt{N}$ 
 -- the center $(0,0)$, corresponding to the relevant
 Dirac point,
is omitted from the sum.  
For $k>1$ the corresponding sums converge which proves
\eqref{Ikest}.\\
For $k=1$ we will establish constant $C_1$ and $C_2$
such that
\begin{equation}
  C_1 \ln N <  \sum_{(p,q)\in L}
  \frac{1}{S_{\underline{K}, 11}p^2 + 2S_{\underline{K}, 12}pq+
    S_{\underline{K}, 22}q^2 } < C_2 \ln N
  \label{logineq}
\end{equation}  
which then directly leads to \eqref{I2est}.
To establish $C_1$ note that because each term 
in the sum \eqref{logineq} is positive 
its value decreases by restricting it
to a square region 
$- a_1 \sqrt{N} \le p \le a_1 \sqrt{N}$,
$- a_1 \sqrt{N} \le q \le a_1 \sqrt{N}$ 
which is completely contained in $L$.
Up to an error of order one 
the sum over a square region can in turn be written 
as a sum over eight terms of the form 
\begin{equation}
  \sum_{p=1}^{a_1 \sqrt{N}} \sum_{q=1}^p 
  \frac{1}{S_{\underline{K}, 11}p^2 \pm 
    2S_{\underline{K}, 12}pq+
    S_{\underline{K}, 22}q^2} \ . 
\end{equation}
For fixed $p$ we can find $q_{\mathrm{max}}$ such that 
\begin{multline}
  \sum_{q=1}^p \frac{1}{S_{\underline{K}, 11}p^2 \pm 
    2S_{\underline{K}, 12}pq+
    S_{\underline{K}, 22}q^2}>\\
  \frac{p}{S_{\underline{K}, 11}p^2 \pm 
    2S_{\underline{K}, 12}pq_{\mathrm{max}}+
    S_{\underline{K}, 22}q_{\mathrm{max}}^2}\ .
\end{multline}
We may choose  $q_{\mathrm{max}}= b_1 p$ for some constant $b_1\ge 0$,
so 
\begin{equation}
  \sum_{p=1}^{a_1 \sqrt{N}} \sum_{q=1}^p \frac{1}{S_{\underline{K}, 11}p^2 \pm 
    2S_{\underline{K}, 12}pq+
    S_{\underline{K}, 22}q^2} >
  c \sum_{p=1}^{\sqrt{N}} \frac{1}{p}
\end{equation}  
which diverges as $\ln N$.

Establishing $C_2$ and the corresponding
logarithmic bound from above follows the same line by first extending
the sum to a square of side length $2 a_2 \sqrt{N}$ 
that completely contains $L$ and then establishing
\begin{multline}
  \sum_{q=1}^p \frac{1}{S_{\underline{K}, 11}p^2 \pm 
    2S_{\underline{K}, 12}pq+
    S_{\underline{K}, 22}q^2}<\\
  \frac{p}{S_{\underline{K}, 11}p^2 \pm 
    2S_{\underline{K}, 12}pq_{\mathrm{min}}+
    S_{\underline{K}, 22}q_{\mathrm{min}}^2}\ 
\end{multline}
with $q_{min}=b_2 p$.\\
Let us note in passing that estimates based
on Poisson summation reveal more detail, i.e.
\begin{align}
  I_2=& A \ln N + O(1), 
  \label{I2est1}\\
  I_{2k}= & 2 \sqrt{3} N^{k-1} \left(Z_2(S_{\underline{K}},k)+ Z_2(S_{\underline{K}'},k)
  \right)+\nonumber\\
  &\quad  + O(N^{k-2})\qquad\qquad \qquad\text{for $k\ge 2$,}
  \label{I2kest1}
\end{align}
where $A>0$ is a constant. A rigorous treatment of the $O(1)$ estimate is more 
involved, however. In fact, \eqref{I2est}and \eqref{Ikest} are sufficient in the context of this paper.

We note that each term in $Z_{2}\left(S_{\underline{K}},k\right)$ is smaller than 
the corresponding term in $Z_{2}\left(S_{\underline{K}},2\right)$ for $k>2$, and so it 
follows that $Z_2(S_{\underline{K}},k)< Z_2(S_{\underline{K}},2)$ for $k>2$.
This property
of the Epstein zeta function
and the estimate \eqref{Ikest} 
imply 
$\sum_{n=2}^\infty I_{2n} E_+^{2n-1}< \frac{C}{N E_+}\sum_{n=2}^\infty (N E_+^2)^n$
i.e. the infinite sum in \eqref{exp}
converges for $E_+< 1/\sqrt{N}$.

Let us now show that one indeed finds a solution of $F(E_+)=0$
using the expansion \eqref{exp} inside the convergence radius.
We start with the estimate that is obtained by truncating
the sum in \eqref{exp} at $n=1$, i.e. $\frac{4 \sqrt{3}}{N E_+}
- I_2 E_+=0$. This estimate gives $E_+^2\approx 
\frac{4 \sqrt{3}}{N I_2}$ -- for sufficiently large
$N$ this is in the radius of convergence of the complete
expansion \eqref{exp}. We will now show rigorously
that this estimate gives the leading order
correctly. 
First of all, the estimate implies that a zero
inside the radius of convergence exists. Moreover since
$I_{2n}>0$ all terms of \eqref{exp} that have been neglected in the 
estimate enter with the same sign. So the true value $E_+>0$ has to
be smaller than the estimate, which we write as
\begin{equation}
  E_+^2=\frac{4 \sqrt{3}}{N I_2}- \Delta >0
\end{equation}
with $\Delta> 0$. We will show rigorously that $\Delta/E_+ \to 0$ as 
$N \to \infty$. Indeed one may rewrite $F(E_+)=0$ in the form
\begin{align}
  &&\frac{4 \sqrt{3}}{N}- I_2 E_+^2 = &\sum_{n=2}^\infty I_{2n} E_+^{2n}
  \qquad \quad \\
  &\Rightarrow&
  I_2 \Delta = &\sum_{n=2}^\infty I_{2n} E_+^{2n} .
\end{align}
Following the same arguments as used above for the calculation
of the convergence radius and using the already established fact that
$E_+$ is inside the convergence radius for sufficiently large $N$
we get the following inequality
\begin{eqnarray*}
    0 < N I_2 \Delta &<& C \sum_{n=2}^\infty 
    (N E_+^2)^n\\
    &=& \frac{C N^2 E_+^4}{1- N E_+^2}= O(I_2^{-2}).
\end{eqnarray*}
So $\Delta = O(I_2^{-3}N^{-1})= O\left((\ln N)^{-3} N^{-1}\right)$, 
or $\Delta/E_+= O\left((\ln N)^{-5/2} N^{-1/2} \right)$. We thus obtain
\begin{equation}
E_+^2 = \frac{4 \sqrt{3}}{N I_2}\left( 1 - O({(\ln N)^{-2}})\right), 
\end{equation}
and analogously
\begin{equation}
  F'\left(E_{\pm}\right) = -2I_{2} + O\left(\frac{1}{\ln N}\right).
\end{equation}

This allows us to show that only two states have an overlap $O(1)$ 
with the starting state and are thus the relevant states to be 
considered in the time-evolution of the algorithm.  
Using the definitions of $\ket{\psi_{a}}$ and $R_{a}$ 
in Eqs.~(\ref{eq:perturbed_eigenvector_definition}) and (\ref{eq:Ra}), 
the inner product of the starting state and the perturbed 
eigenvectors can be expressed as 
\begin{equation}
  \braket{start}{\psi_{a}} = \frac{3^{\frac{1}{4}}}{E_{a} |F'\left(E_{a}\right)|^{\frac{1}{2}}}\braket{start}{\alpha_{o},\beta_{o}}.
\end{equation}
Applying our previous results for the states nearest to the Dirac point, we see that
\begin{equation}
  |\braket{start}{\psi_{\pm}}| \approx \frac{1}{\sqrt{2}} + O\left(\frac{1}{\ln^{2} N}\right).
\end{equation}
Our starting state is thus a superposition of the perturbed eigenstates, $\ket{\psi_{\pm}}$, which also facilitate the search algorithm, see the main text. This now allows us to investigate the running time and success amplitude of the algorithm by looking at the time-evolution, that is,
\begin{align}
&\left|\bra{\ell}e^{-iHt}\ket{start}\right|  \\
&\approx \left|\frac{1}{\sqrt{2}}\left(e^{-i E_{+}t}\braket{\ell}{\psi_{+}} - e^{i E_{+}t}\braket{\ell}{\psi_{-}}\right)\right| \\
&= \frac{1}{3^{\frac{1}{4}}I_{2}^{\frac{1}{2}}}\left|\sin\left(E_{+}t\right)\right|
\end{align}
It is clear from our earlier results for $E_{\pm}$ and $I_{2}$ that our algorithm localizes on the neighbor state 
$\ket{\ell}$ in time $T = \frac{\pi}{2 E_{+}} = O\left(\sqrt{N\ln N}\right)$ with probability amplitude $O\left(1/\sqrt{\ln N}\right)$.\\[1cm]

\noindent
[S1] {\em Spatial search by quantum walk}, Andrew M. Childs and Jeffrey Goldstone, Phys. Rev. A. 70:022314 (2004).

\noindent
[S2] {\em On Advanced Analytic Number Theory}, C.L. Siegel, Tata Institute of Fundamental Research (1961).

\end{document}